\documentclass[twocolumn,superscriptaddress,pra]{revtex4-1}
\usepackage{amsfonts}
\usepackage{amsmath}
\usepackage{graphicx}
\usepackage{hyperref}
\usepackage{bbm}
\usepackage{color}
\usepackage{notes2bib}

\begin{document}
\title{Combining the synergistic control capabilities of modelling and experiments: illustration of finding a minimum time quantum objective}
\author{Qi-Ming~Chen*}
\affiliation{Department of Chemistry, Princeton University, Princeton, New Jersey 08544, USA}

\author{Xiaodong~Yang*}
\affiliation{Department of Chemistry, Princeton University, Princeton, New Jersey 08544, USA}
\affiliation{Hefei National Laboratory for Physical Sciences at the Microscale and Department of Modern Physics, University of Science and Technology of China, Hefei 230026, China}
\affiliation{CAS Key Laboratory of Microscale Magnetic Resonance, University of Science and Technology of China, Hefei 230026, China}

\author{Christian~Arenz*}
\affiliation{Department of Chemistry, Princeton University, Princeton, New Jersey 08544, USA}

\author{Re-Bing~Wu}
\affiliation{Department of Automation, Tsinghua University \emph{\&} Center for Quantum Information Science and Technology, BNRist, Beijing 100084, China}

\author{Xinhua~Peng}
\affiliation{Hefei National Laboratory for Physical Sciences at the Microscale and Department of Modern Physics, University of Science and Technology of China, Hefei 230026, China}
\affiliation{CAS Key Laboratory of Microscale Magnetic Resonance, University of Science and Technology of China, Hefei 230026, China}
\affiliation{Synergetic Innovation Centre of Quantum Information \emph{\&} Quantum Physics, University of Science and Technology of China, Hefei, Anhui 230026, China}

\author{Istv\'an~Pelczer}
\affiliation{Department of Chemistry, Princeton University, Princeton, New Jersey 08544, USA}

\author{Herschel~Rabitz}
\affiliation{Department of Chemistry, Princeton University, Princeton, New Jersey 08544, USA}


\begin{abstract}
A common way to manipulate a quantum system, for example spins or artificial atoms, is to use properly tailored control pulses. In order to accomplish quantum information tasks before coherence is lost, it is crucial to implement the control in the shortest possible time. Here we report the near time-optimal preparation of a Bell state with fidelity higher than {$99\%$} in an NMR experiment, which is feasible by combining the synergistic capabilities of modelling and experiments operating in tandem. The pulses preparing the Bell state are found by experiments that are recursively assisted with a gradient-based optimization algorithm working with a model. Thus, we exploit the interplay between model-based numerical optimal design and experimental-based learning control. Utilizing the balanced synergism between the dual approaches, as dictated by the case specific capabilities of each approach, should have broad applications for accelerating the search for optimal quantum controls.  
\end{abstract}
\maketitle

\section{Introduction}

The precise dynamical manipulation of a quantum system in a time-optimal manner is crucial for constructing high-fidelity quantum devices \cite{Rabitz2000}. In particular, in order to operate on a timescale faster than the shortest coherence time, the creation of entangled states in the most concise time is of high relevance for quantum information science \cite{Bennett2000}. Here we demonstrate the preparation of the Bell state $|\psi_g\rangle=\frac{1}{\sqrt{2}}\left( |10\rangle - |01\rangle \right)$ resulting in fidelity higher than {$99\%$} while also reaching close to the shortest possible preparation time $T_{\rm min}$ \cite{d2007introduction}. We achieve this performance by combining an operationally guided balance of closed-loop learning experiments with model-based numerical design, which allows for the correction of systematic errors caused by possible uncertainties in the model and reduce the costly experimental tomography at each step. This balanced approach enhances the efficiency of finding optimal controls while also assuring quality dual objective performance in the present illustration.

\begin{figure} [h!]
  \centering
  \includegraphics[width=8cm]{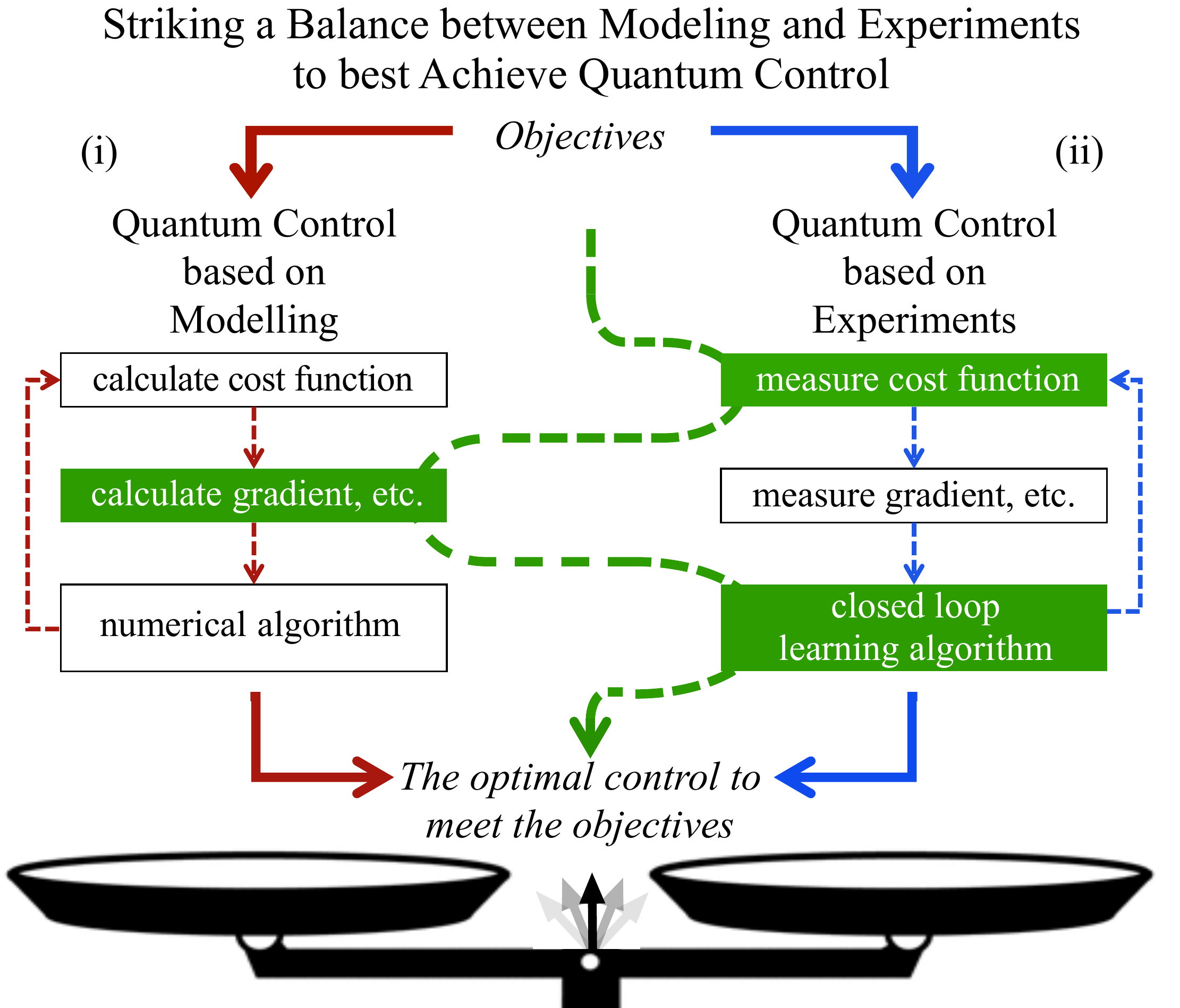}
  \caption{\label{fig:scheme} (Color online) Illustration of two general optimization approaches to find the controls achieving desired objective(s). In (i) the optimization loop is entirely based on numerical simulation-based design (red) followed by implementation of the design, whereas (ii) makes exclusive use of experiments and data analysis (blue). The green dashed curve indicates a balanced synergistic combination of both approaches, which is used in our NMR experiment to (a) prepare a Bell state while (b) also in minimum time. In this illustration the controls are updated using a gradient-based closed-loop learning algorithm, in which we utilize a model-based simulation to calculate the gradient and then experimental state tomography to determine the step size in each iteration cycle. In general, the best balanced way to combine the two approaches (i) and (ii) depends on the experimental platform considered, the quality of the model describing the system, the objectives, and the algorithm(s) used to achieve the desired task. Thus, the balance between (i) and (ii) is not static, as it rests on the future progress made in technology as well as theory and computational capabilities.}
\end{figure}

As indicated above, time-optimal state preparation can be viewed as a dual objective optimization problem in which the norm of the overlap with the target state is maximized utilizing the control fields while their pulse length is minimized. Experimentally realizable solutions to this problem are particularly challenging to find, since high fidelity is aimed for, and the control fields preparing the desired state are asked to be as short as possible while being implementable in the laboratory. In particular, at one extreme (i) solely \emph{model- based} optimization schemes can suffer from uncertainties in the underlying model diminishing the fidelity (i.e., including possibly all measures of objective performance) and the designed fields may not be precisely created by the apparatus upon laboratory implementation. At the other extreme (ii) solely \emph{experiment-based} schemes (i.e., often called either learning control or adaptive feedback control) can require an extensive laboratory overhead in some applications. Before we address these challenges, as schematically represented in Fig. 1, we first outline in Sec. \ref{sec:quantumcontrol} the two distinct approaches (i) and (ii) for generally solving quantum control problems. We discuss advantages and disadvantages of each approach, followed by introducing  so called \emph{hybrid approaches} that aim to best overcome the detrimental disadvantages by combining certain aspects of the model-based and the experiment-based optimization schemes. As a challenging illustration, the following material will explain how we combine model-based and experiment-based optimal control (green dashed line in Fig. 1) to prepare the target Bell state with high fidelity in a time close to $T_{\text{min}}$, with details given in Sec. \ref{sec:theory} and Sec. \ref{sec:BellStatepre}. 

\section{Approaches to achieving optimal control objectives}
\label{sec:quantumcontrol}  
  A typical quantum control problem is to find the classical fields that minimizes/maximizes a given cost functional (i.e., the overlap (distance) between the desired state (unitary transformation) and the achieved state (evolution) of the system). The cost functional is often subject to additional demands such as minimizing the energy or the length of the pulses, which is considered in this work.  As mentioned in the Introduction, there are generally two approaches (i) and (ii) to solve such quantum control problems.      

\subsection{Quantum Control based on Modelling (i)}\label{sec:QCmodelling}
 In the first approach (i) in Fig. 1 the optimization is performed on a classical computer using standard iterative procedures, such as gradient or stochastic algorithms, that find the fields maximizing/minimizing the desired objectives \cite{Peirce1988, Khaneja2005, Dolde2014, Glaser2015, Wang2015, Geng2016}; the resultant fields are implemented in the laboratory in a final single step. However, this approach requires highly reliable knowledge of the model describing the system. Any erroneous system parameters as well as other significant missing (possibly unwittingly so) model components will diminish the utility of the designed pulses resulting in the quality of the achieved tasks likely dropping in the final experimental performance test. Furthermore, implementation of the control design in the laboratory apparatus can also introduce undesirable distortions in the actually created control.

\subsection{Quantum Control based on Experiments (ii)}\label{sec:QCexperiment}
In contrast, in the second approach (ii) the classical fields achieving the objectives are directly "learned" in experiments \cite{Judson1992, Phan1999, Warren1993, Lee2002, Sun2014, Sun2015, Sun2017, Li2017, Lu2017}. This procedure has the advantages that (a) no detailed prior knowledge of the system is required, (b) the physically exact scenario is employed with all parameters at their true values, (c) the possibly numerically expensive and/or error prone calculations with the time-dependent Schr\"odinger equation are avoided and instead directly performed in an analog fashion by the real system in the experiment where due care is needed to deal with noise from various sources. Here optimization is fully dependent on repeated measurements of system observables and the control fields are updated accordingly. For instance, control of a multi-qubit system was successfully demonstrated \cite{Li2017, Lu2017} by iteratively measuring the gradient of the objective with respect to the controls and appropriately updating the controls. The pulses were found to give more accurate results than those generated by model-based numerical optimization, showing that a measurement-based optimization procedure can correct for unknown systematic imperfections. However, depending on the platform, the objectives and the optimization algorithm used, such measurement-based strategies alone (i.e., pathway (ii) in Fig. 1) can become experimentally intensive under various circumstance with current technology. As an example, for high-quality time optimal control implementations, a large number of tomography experiments are required since the gradient with respect to the controls as well as the gradient with respect to the evolution time needs to be measured (see Sec. \ref{ref:expbased}). 

\subsection{Hybrid Quantum Control approaches}  

In order to overcome the aforementioned issues of each approach (i) and (ii), combinations of both approaches have been considered \cite{egger2014adaptive, dive2017situ, wu2018data}. For instance, in \cite{egger2014adaptive} (i) a gradient algorithm based on a model was used to first achieve \emph{near-optimal} solutions, followed by (ii) experimentally optimizing the obtained pulses through tomography measurements further to obtain even higher fidelities.  Instead of starting with approach (i) and continuing with approach (ii), we address the conundrum of operating with either (i) or (ii) above by combining both approaches in a \emph{balanced fashion}, dictated by the particular circumstances. As schematically represented in Fig.~1 (green dashed curve), in the present NMR case instead of measuring the gradient in each iteration step, we numerically \emph{calculate} the gradient based on a model. The controls are updated with a step size that depends on the fidelity for preparing the target state, which is \emph{measured} in each iteration step. This procedure has the advantage that even if the adopted model is inaccurate, in each iteration step it is ensured that the objective is more closely approached as long as the gradient points in the "climbing" direction. This approach allows for uncertainties in the model causing even a moderate level of systematic error and avoids an otherwise unacceptable number of gradient measurements, as would happen when operating on path (ii) alone in the present experiment. We remark that the balanced approach in Fig. 1 should be viewed with application specific freedom, including the prospect of (i) and (ii) each utilizing different types of algorithms with the criteria of maximally useful information exchange between (i) and (ii) leading to increased efficiency and better final objective(s) performance. The realization of various forms of the algorithmic freedom offered in Fig. 1 must await future research, but this paper will give a specific illustration. In this regard path (i) can exploit access to the evolving state, while path (ii) nominally only has access to information at time $T$. 

\section{Description of the illustrative system}
Our aim is demonstrate the balanced approach to find the control pulses that prepare the target Bell state $|\psi_g\rangle$ to high fidelity in the shortest attainable pulse time $T_{\text{min}}$. Before we define the corresponding quantum control problem in Sec. \ref{sec:theory}, we first describe the experimental setting as well as the model describing the control system.   

The experiment is performed on a Bruker  Avance III HD $800\,\rm{MHz}$ spectrometer at temperature $295\,\rm{K}$, in which the states of the two spins $\rm {}^{13}C$ and $\rm {}^{1}H$ are encoded as the two qubits in the labeled chloroform sample ($^{13}\rm CHCl_{3}$) dissolved in $\rm DMSO-d_{6}$. The relaxation times $T_1$ and $T_2^{*}$ are measured to be $T_{1}=730\,\text{ms},~T_{2}^{*}=96.5\,\rm{ms}$ for  $\rm {}^{13}C$ and $T_{1}=96\,\text{ms},~T_{2}^{*}=42.5\,\text{ms}$ for $\rm {}^{1}H$, respectively. In the rotating frame, the interaction between the two spins is described by the drift Hamiltonian
\begin{align}
\label{eq:driftH}
H_0=\frac{\pi}{2} g \sigma_z^{1}\sigma_z^{2},
\end{align}
 where $\sigma_{j}^{1,2}$ with $j=x,y,z$ denotes the Pauli operators on the two spins, and the coupling constant is measured to be $g=217.4\,\rm{Hz}$. The external controls $\{u_{j}^{k}(t)\}$ applied on the system are included in the time-dependent control Hamiltonian
 \begin{align}
 \label{eq:controlH}
  H_{c}(t)=\sum_{k=1,2} \pi[u_x^{k}(t) \sigma_x^{k}+u_y^{k}(t)\sigma_{y}^{k}]. 
  \end{align}
 The total Hamiltonian is then given by $H(t)=H_{0}+H_{c}(t)$. In the experiment, we initially prepare the system in the state $|\psi(0)\rangle=|00\rangle$  with fidelity $0.999$  \bibnote{In the NMR community this result is sometimes referred to as a pseudo-pure state} determined by using the line-selective method \cite{Peng2001}. 

\section{The Time-Optimal Control Problem}\label{sec:theory}
Time-optimal state preparation can be formulated as the dual-objective optimization problem 
\begin{align}
\label{eq:optimizationproblem}
	&\max_{\{u_{j}^{k}(t)\}}~~\left [ J(\{u_j^k(t)\}, T) \right],\nonumber \\	
	&~\text{while~minimizing~}T,
\end{align}
subject to satisfying the Schr\"odinger equation, where $J(\{u_j^k(t)\}, T)=|\langle \psi_{\text{g}}|\psi(T)\rangle|^{2}$ is the fidelity, $|\psi(T)\rangle=U(T)|\psi(0)\rangle$ is the state of the system at  $T>0$ and $|\psi_g\rangle=\frac{1}{\sqrt{2}}\left( |10\rangle - |01\rangle \right)$ is the desired target Bell state. The control pulses  $\{u_{j}^{k}\}$ of length $T$ enter in the time evolution operator $U(T)$ through the control Hamiltonian \eqref{eq:controlH}. In the experiments we assume that the controls are piecewise constant over $M=50$ uniform intervals and the $m$-th control amplitude of the control $u_{j}^{k}(t)$ is denoted by $u_{j}^{k}[m]$. 

Assuming the control fields are unconstrained, we proceed by determining the smallest possible time $T=T_{\text{min}}$ at which the target Bell state can be prepared with fidelity 1, followed by introducing the algorithm that is employed in this work to solve \eqref{eq:optimizationproblem}.

\subsection{Theoretical value for the minimum time}
In general, finding an analytical expression for the minimum time $T_{\text{min}}$ (e.g., to implement a unitary gate or prepare a state in a generic quantum system) remains an unsolved problem. Although some progress has recently been made by developing an upper bound on $T_{\text{min}}$ for qubit networks \cite{Arenz2018}, the exact value is only known for low dimensional systems \cite{Hegerfeldt2013, Khaneja2001, Khaneja2002, Carlini2006}. In seminal work \cite{Khaneja2001} the minimum time for implementing a generic unitary transformation on a two-spin system was determined; this information will be used as a comparative benchmark in the present work. We begin by noting that every unitary operation on a two-spin system can be decomposed as
\begin{align}
\label{eq:optimalpulse}
	U=V\exp[-i(a_{x}\sigma_{x}^{1}\sigma_{x}^{2}+a_{y}\sigma_{y}^{1}\sigma_{y}^{2}+a_{z}\sigma_{z}^{1}\sigma_{z}^{2})]W,
\end{align}
where $V$ and $W$ are local unitary operations in  $\text{SU}(2)\otimes \text{SU}(2)$ \cite{Khaneja2001}. In the case where the strength of the control Hamiltonians can be made arbitrarily large, every local operation on each spin can be created instantaneously. The minimum time $T_{\text{min}}$ to produce $U$ is then determined by the smallest value of $\sum_{j}|a_{j}|$. Starting from the initial state $|00\rangle$, it is easy to find local rotations $V$ and $W$ that prepare, for $a_{x}=a_{y}=0$, the target Bell state $|\psi_g\rangle$. Assuming that the control fields are unconstrained, we find \cite{Khaneja2001} that the minimum time to prepare $|\psi_g\rangle$ is given by $T_{\text{min}}={1}/{(2g)}=2.30\,\rm{ms}$, which is significantly below the relaxation times $T_{1}$ and $T_{2}^{*}$. Clearly, the assumption of infinitely strong control fields is unphysical in practice. However, as we will show in Sec \ref{sec:balanceAP} (Fig. 2 (a) right panel), using the synergistic balanced optimization approach in Fig. 1 yields smooth pulses that prepare the target Bell state with high fidelity very close to $T_{\text{min}}$.

\subsection{Algorithm}\label{sec:algorithm}

To solve \eqref{eq:optimizationproblem} we resort to the numerical method which is detailed in \cite{Chen2015}. In the following we describe the algorithm for time optimal state preparation, which in the present application is independent of whether approach (i) (Sec. \ref{sec:QCmodelling}), or (ii) (Sec. \ref{sec:QCexperiment}) is used as well as a combination of both indicated in Fig. 1. In the next section, we will explain how we combined the approaches (i) and (ii) in the experiments to prepare the target Bell state close to the minimum time, while taking advantage of the distinct efficiencies and circumstances along paths (i) and (ii). 

The first step is to choose a pulse length $T$ sufficiently large so that a high fidelity can be achieved and then we employ a standard gradient algorithm to find the controls that reach a prescribed fidelity $J_{H}$ where $H$ refers to the "highest" attainable value. For a fixed $T$, the control variables are updated in the direction of the gradient of the fidelity with respect to the piecewise constant controls. Thus, in each iteration step a change in $u_{j}^k[m]$ reads  $\Delta u_{j}^k[m] = g_{u_{j}^k[m]}$ where $g_{u_{j}^k[m]}$ is the gradient with respect to the $m$-th control value.  An update of the controls is accepted if the inequality,  	
\begin{align}
\label{eq:ineq1}
J(\{u_j^k[m]+d_{1}\Delta u_j^k[m]\}, T) &\geq J(\{u_j^k[m]\}, T) \nonumber \\ 
  &+\alpha d_{1}\sum_{j,k,m}\Delta u_j^k[m] g_{u_j^k[m]},
\end{align}
is satisfied, with $\alpha=0.01$ being a constant which enables acceptable convergence efficiency
\cite{boyd2004convex} and $d_{1}$ is the step size in step 1. 

In a second step, both the control variables as well as the pulse length are simultaneously changed according to $\Delta u_{j}^k[m] = g_{u_{j}^k[m]}/g_T$ and $\Delta T = -\sum_{j, k,m} \left( \Delta u_{j}^k[m] \right)^2$, where $g_{T}$ is the gradient with respect to the pulse length $T$, while aiming to keep the achieved control fidelity $J_{H}$ unchanged. An update is accepted if the inequality 
\begin{align}
\label{eq:ineq2}
J(\{u_j^k[m]+d_{2}\Delta u_j^k[m]\}, T+d_{2}\Delta T) \geq \beta J(\{u_j^k[m]\}, T),
\end{align}
 is satisfied, where $\beta=0.999$ manages the rate of deviation from $J_{H}$ and $d_{2}$ denotes the step size used in step 2. If the fidelity decreases to a lower threshold value $J_{L}$ due to numerical or experimental errors, then we return to the first step using the current $T$, which remains fixed in order to climb to $J_{H}$ again. This procedure is repeated until the target state is reached with high fidelity $J_{H}$ while reaching the smallest attainable final time \bibnote{Since $J_{H}=0.999$ is fixed we remark that the present procedure seeks a single point along the Pareto front.}.

\section{Bell state preparation close to the minimum time}
\label{sec:BellStatepre}

 \begin{figure*}
  \centering
  \includegraphics[width=18cm]{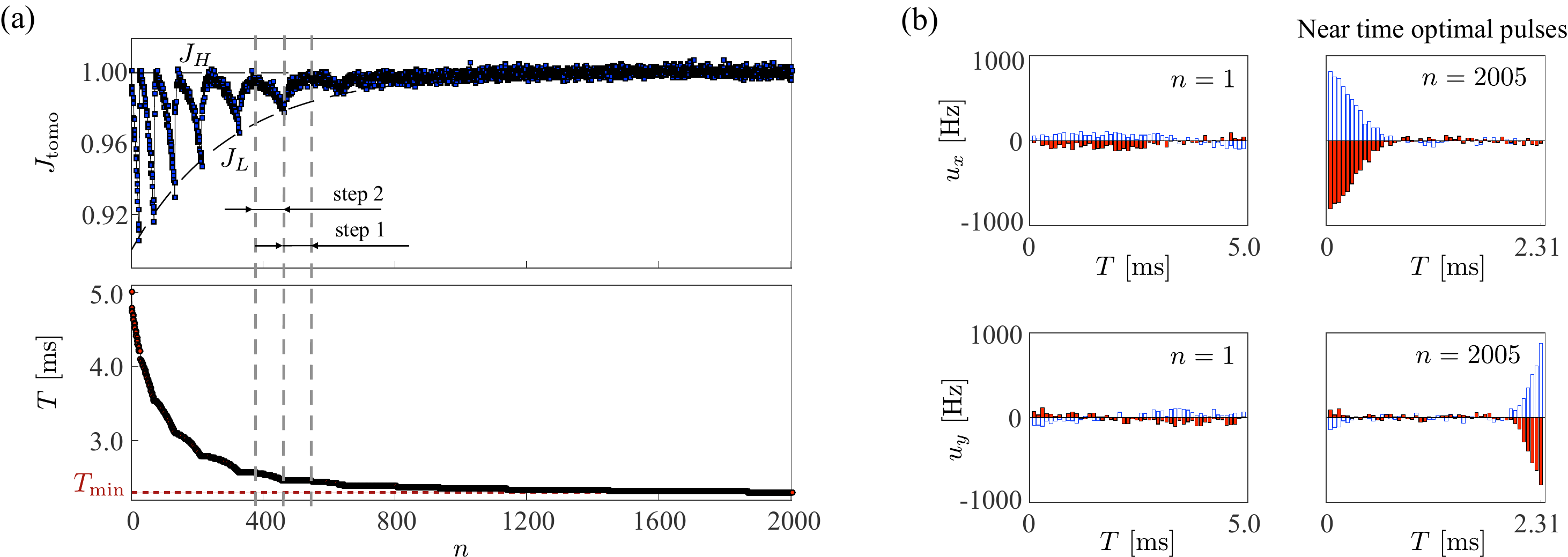}
  \caption{\label{fig:data} (Color online) Experimental data for preparing a Bell state with high fidelity near the minimum time $T_{\text{min}}=2.30\,\rm{ms}$ (dashed red line) using shaped control pulses obtained by employing a synergistic combination of approaches (i) and (ii) shown in Figure~\ref{fig:scheme}. (a) shows the partial fidelity $J_{\rm tomo}$ obtained from 3 measurements on spin 1 and the time length $T$ of the control pulses as a function of the iteration step $n$ of the optimization algorithm. The curved black dashed line is the lower threshold $J_{L}$ and the straight horizontal black dashed line is the prescribed fidelity $J_{H}$. The vertical grey dashed lines indicate the two steps of the algorithm that was introduced in section \ref{sec:algorithm}, including a fidelity climb back to $J_{\text{H}}$ when a reduction to the threshold $J_{\text{L}}$ occurs. As explained below Eq. \eqref{eq:ineq2}, when $J_{\rm tomo}$ drops beyond $J_{L}$ due to experimental or modelling errors, $T$ is kept fixed to bring the fidelity back to $J_{H}$ (i.e., indicated by the fidelity reclimbing in the early stages of the procedure), followed by further shrinking of $T$.  (b) shows the optimized control pulses applied on spin 1 (blue open bars) and spin 2 (red solid bars) for a pulse length of $T=5.00\,\rm{ms}$ (left panel) and the pulses of length $T=2.31\,\rm{ms}$ near the minimum time (right panel).} 
\end{figure*}

We now turn to applying the algorithm introduced in Section \ref{sec:algorithm} to solve \eqref{eq:optimizationproblem}. As outlined in Sec. \ref{sec:quantumcontrol}, there are two distinct approaches (i) and (ii) to solve quantum controls problems. Based on the algorithm introduced in Sec. \ref{sec:algorithm}, we will now describe in more detail the two approaches (i) and (ii) for finding the control fields that prepare the target Bell state close to the minimum time, followed by combining the best features (i) and (ii) in a balanced fashion to overcome the issues of each approach.

\subsection{Solely model-based design (i)}\label{ref:modelbased}
Assuming the model introduced in \eqref{eq:driftH} and \eqref{eq:controlH} is a high quality description of the actual NMR system, in the present case perhaps the simplest approach is to run the algorithm entirely on path (i) in Fig. 1 and then implement the resultant optimal pulses in the laboratory. Following this logic and starting from randomly chosen initial pulses we reached a simulation fidelity of $J_{H}=0.999$ with highly peaked pulses (not shown here), yielding a minimum time of $T_{\text{min}}=2.27\,\rm{ms}$. We note that this time is below the theoretical value $T_{\text{min}}=2.30\,\rm{ms}$, which is, as described in the previous section, obtained for delta function like pulses and a \emph{perfectly} prepared target Bell state, i.e., with fidelity 1. Implementing the numerically obtained pulses in the laboratory produces a fidelity of $J=0.976$ for preparing the target Bell state, which was measured using full state tomography. We note that the statistical error of the $800\,\rm{MHz}$ NMR machine is of the order $\sim 10^{-4}$, which is confirmed by repeating the same control and measurement pulses for $5$ times. The partial degradation of the fidelity arises from the fact that large scale sudden jumps between power levels in the designed pulses can be somewhat "distorted" while utilizing high Q cryoprobes with high sensitivity, as the changes in the pulses are faster, or correspond to higher frequencies, than the bandwidth of the resonant circuit \cite{PrivateComBruk}. There are also additional experimental factors, including low level instrument and spin/sample noise, small local temperature and homogeneity fluctuations, finite relaxation times and cross correlation relaxation, intra- and intermolecular nuclear Overhauser effects and multiple quantum effects, etc. \cite{sorensen2006james}, that cannot be readily accounted for by the designed pulses.

\subsection{Solely experiment-based design (ii)}\label{ref:expbased}

Turning to the other side of the balance, i.e., using just approach (ii) that does not rely on a model, ensures that the optimization procedure is performed within the capabilities of the experimental apparatus (i.e., see remarks at the end of the last paragraph). However, as mentioned earlier, using just (ii) with measurement data and closed-loop learning to solve the present time-optimal state preparation problem directly in the laboratory requires an impractically large number of measurements, which can be estimated as follows. We first note that, due to the symmetry of the \emph{target} Bell state, rather than performing  full tomography of the state, the fidelity can be read out through 3 measurements only on the 1st spin \cite{Vandersypen2005}.  We denote the fidelity measured with partial tomography as $J_{\rm tomo}$ to distinguish it from $J$. We note that, since the state obtained experimentally during the optimizationt does not necessarily possess this symmetry, the actual fidelity $J$ inferred by full tomography can be different. However, measuring $J_{\text{tomo}}$ and the gradient of the four control fields $\{ g_{u_{j}^k[m]}\}$ requires $3$ and $4\times 50\times 2 \times 3$ independent measurements \cite{Li2017}, respectively. The number $3$ comes from the partial tomography in 3 directions on the 1st spin. Furthermore, the number of measurements required to obtain $g_{T}$ needed in the second step of the algorithm is $50\times 2 \times 3$. Thus, assuming the number of iterations is $n\sim 2000$, the total experimental time is about $7500\,\rm{h}$ based on the fact that each measurement takes $10\,\rm{s}$. This time duration makes the fully experimental method, i.e., path (ii) in Fig. \ref{fig:scheme}, at best difficult to be employed in solving time-optimal control problems.

\subsection{Balanced approach}\label{sec:balanceAP}

Based on the considerations in Section \ref{ref:modelbased} and Section \ref{ref:expbased},  we adopt the combined procedure in Fig. 1, and swing back and forth between the approaches (i) and (ii) drawing on the best features of both. This process entails the steps:
\begin{itemize}
\item [(a)] in each iteration we \emph{measure} the fidelity $J_{\text{tomo}}$ using partial state tomography,
\item [(b)] the gradient with respect to the length of the pulse $g_{T}$ and the gradients with respect to the control field amplitudes $\{g_{u_{j}^{k}[m]}\}$ are \emph{calculated} numerically during the experiment based on the model given in \eqref{eq:driftH} and \eqref{eq:controlH},

\item[(c)] the step sizes $d_{1}$ and $d_{2}$ are adjusted in each iteration according to the measured (partial) fidelity $J_{\text{tomo}}$,

\item[(d)] the controls are updated when \eqref{eq:ineq1} and \eqref{eq:ineq2} are satisfied. 
\end{itemize}
At the end of the iteration sequence we determine how well the target Bell state was prepared by performing full tomography to obtain $J$. We set the lower threshold to be $J_L=0.999 - 0.099 e^{-n/300}$ to indicate our tolerance for observed fidelity loss when changing $T$, which converges to $J_H=0.999$ when the iteration number $n$ gets large (i.e., assuming that experimental artifacts do not forbid reaching $J_{H}$). When operating above the lower threshold $J_{L}$, the step size is shrunk according to our measurements in the laboratory.

Since only the fidelity $J_{\text{tomo}}$ is measured in each iteration and not the gradient, the total experimental time with the balanced approach can be decreased to $16.7\,\rm{h}$ which allows for reducing the number of measurements required by more than $\sim$ two orders of magnitude.

The iterative process of the algorithm in the NMR experiment is shown in Fig.~2~a) with details found in the caption. The partial fidelity $J_{\rm tomo}$ (upper panel) for preparing the target Bell state as well as the length of the control pulses $T$ (lower panel) is plotted as a function of the iteration step $n$.  Starting from randomly chosen initial pulses, in \emph{step 1} the target Bell state was prepared with fidelity $J=0.999$ ($J_{\rm tomo}=0.995$) by optimizing the controls at setting $T=5.00\,\rm{ms}$. The corresponding optimized control pulses are shown in the left panel of Fig.~2~b), wherein red (blue) corresponds to the control pulse on spin 1 (2); the number of piecewise constant values $M=50$ used for the controls in the laboratory are the same as in the simulations on path (i), as stated earlier. According to the optimization algorithm described in the last paragraph, the total evolution time is iteratively reduced while the controls are optimized in each step. After $n=2005$ iteration steps the length of the pulses is reduced to $T = 2.31{\rm ms}$, which is close to the theoretical minimum time $T_{\text{min}}=2.30\,\rm{ms}$, shown as a red dashed horizontal line in Fig. 1 (a). At $T=2.31\,\rm{ms}$ a partial fidelity of $J_{\rm tomo}=0.999$ is obtained. Full tomography of the final state confirms that a Bell state with fidelity $J=0.991$ was prepared at the time $2.31\,\rm{ms}$. The corresponding optimal control pulses are shown in the right panel of Fig.~2~b).
~
\\

\section{Conclusions and Outlook}
We have demonstrated the near time-optimal preparation of a Bell state with high fidelity using shaped (control) pulses in an NMR experiment. To the best of our knowledge, this experiment is the first demonstration of the preparation of a high-fidelity two-qubit entangled state close to the shortest possible time needed for its preparation. The pulses achieving this goal were obtained using a gradient based closed-loop learning algorithm carried out with a synergistic combination of measurement results of the fidelity and model-based numerical calculations of the gradients. This operational balanced combination allows for systematic uncertainties in the model to be corrected in an iterative fashion with the experiments, thereby drawing together the special capabilities of modelling and experiments, as shown in Fig. 1, in order to accurately control quantum systems. Especially for quantum information tasks that rely on high accuracy, a suitable combination of the two optimization approaches (i) and (ii) can open up a practical new way towards achieving high fidelity and robust quantum information processing. The "best" balance for combining the two approaches depends on the capabilities of the particular experimental platform considered (i.e., including various noise sources and their impacting factors), the quality of the model describing the system and its associated computational effort, the objectives, and the algorithm(s) used to achieve the desired task. Since experimental technology and theory/simulation tools naturally continue to develop over time, the optimal implementation of the  balance in Fig. 1 is not static. A case-by-case decision needs to be made about the best way to use the capabilities of models operating with experiments to achieve robust, accurate and scalable implementations.

\section*{Acknowledgements.}
Q.M.C thanks Guilu Long and Tao Xin for their selfless sharing and teaching of NMR techniques and acknowledges partial funding from the DOE (grant DE-FG02-02ER15344). C.A. and H. R. respectively acknowledge funding from the NSF (grant CHE-1763198) and the ARO (grant W911NF-16-1-0014). X. Peng and X. Yang acknowledge fundings from National Key Research and Development Program of China (Grant No. 2018YFA0306600), the National Science Fund for Distinguished Young Scholars (Grant No. 11425523), Projects of International Cooperation and Exchanges NSFC (Grant No. 11661161018), Anhui Initiative in Quantum Information Technologies (Grant No. AHY050000). R.B.W. acknowledges supports from the National Key R\&D Program of China through Grant Nos. 2017YFA0304300 and 2018YFA0306703, and NSFC Grants Nos. 61833010 and 61773232

\bibliography{source19022020}

\begin{thebibliography}{34}%
\makeatletter
\providecommand \@ifxundefined [1]{%
 \@ifx{#1\undefined}
}%
\providecommand \@ifnum [1]{%
 \ifnum #1\expandafter \@firstoftwo
 \else \expandafter \@secondoftwo
 \fi
}%
\providecommand \@ifx [1]{%
 \ifx #1\expandafter \@firstoftwo
 \else \expandafter \@secondoftwo
 \fi
}%
\providecommand \natexlab [1]{#1}%
\providecommand \enquote  [1]{``#1''}%
\providecommand \bibnamefont  [1]{#1}%
\providecommand \bibfnamefont [1]{#1}%
\providecommand \citenamefont [1]{#1}%
\providecommand \href@noop [0]{\@secondoftwo}%
\providecommand \href [0]{\begingroup \@sanitize@url \@href}%
\providecommand \@href[1]{\@@startlink{#1}\@@href}%
\providecommand \@@href[1]{\endgroup#1\@@endlink}%
\providecommand \@sanitize@url [0]{\catcode `\\12\catcode `\$12\catcode
  `\&12\catcode `\#12\catcode `\^12\catcode `\_12\catcode `\%12\relax}%
\providecommand \@@startlink[1]{}%
\providecommand \@@endlink[0]{}%
\providecommand \url  [0]{\begingroup\@sanitize@url \@url }%
\providecommand \@url [1]{\endgroup\@href {#1}{\urlprefix }}%
\providecommand \urlprefix  [0]{URL }%
\providecommand \Eprint [0]{\href }%
\providecommand \doibase [0]{http://dx.doi.org/}%
\providecommand \selectlanguage [0]{\@gobble}%
\providecommand \bibinfo  [0]{\@secondoftwo}%
\providecommand \bibfield  [0]{\@secondoftwo}%
\providecommand \translation [1]{[#1]}%
\providecommand \BibitemOpen [0]{}%
\providecommand \bibitemStop [0]{}%
\providecommand \bibitemNoStop [0]{.\EOS\space}%
\providecommand \EOS [0]{\spacefactor3000\relax}%
\providecommand \BibitemShut  [1]{\csname bibitem#1\endcsname}%
\let\auto@bib@innerbib\@empty
\bibitem [{\citenamefont {Rabitz}\ \emph {et~al.}(2000)\citenamefont {Rabitz},
  \citenamefont {de~Vivie-Riedle}, \citenamefont {Motzkus},\ and\ \citenamefont
  {Kompa}}]{Rabitz2000}%
  \BibitemOpen
  \bibfield  {author} {\bibinfo {author} {\bibfnamefont {H.}~\bibnamefont
  {Rabitz}}, \bibinfo {author} {\bibfnamefont {R.}~\bibnamefont
  {de~Vivie-Riedle}}, \bibinfo {author} {\bibfnamefont {M.}~\bibnamefont
  {Motzkus}}, \ and\ \bibinfo {author} {\bibfnamefont {K.}~\bibnamefont
  {Kompa}},\ }\href {http://science.sciencemag.org/content/288/5467/824}
  {\bibfield  {journal} {\bibinfo  {journal} {Science}\ }\textbf {\bibinfo
  {volume} {288}},\ \bibinfo {pages} {824} (\bibinfo {year}
  {2000})}\BibitemShut {NoStop}%
\bibitem [{\citenamefont {Bennett}\ and\ \citenamefont
  {DiVincenzo}(2000)}]{Bennett2000}%
  \BibitemOpen
  \bibfield  {author} {\bibinfo {author} {\bibfnamefont {C.~H.}\ \bibnamefont
  {Bennett}}\ and\ \bibinfo {author} {\bibfnamefont {D.~P.}\ \bibnamefont
  {DiVincenzo}},\ }\href {\doibase 10.1038/35005001} {\bibfield  {journal}
  {\bibinfo  {journal} {Nature}\ }\textbf {\bibinfo {volume} {404}},\ \bibinfo
  {pages} {247} (\bibinfo {year} {2000})}\BibitemShut {NoStop}%
\bibitem [{\citenamefont {d'Alessandro}(2007)}]{d2007introduction}%
  \BibitemOpen
  \bibfield  {author} {\bibinfo {author} {\bibfnamefont {D.}~\bibnamefont
  {d'Alessandro}},\ }\href@noop {} {\emph {\bibinfo {title} {Introduction to
  quantum control and dynamics}}}\ (\bibinfo  {publisher} {Chapman and
  Hall/CRC},\ \bibinfo {year} {2007})\BibitemShut {NoStop}%
\bibitem [{\citenamefont {Peirce}\ \emph {et~al.}(1988)\citenamefont {Peirce},
  \citenamefont {Dahleh},\ and\ \citenamefont {Rabitz}}]{Peirce1988}%
  \BibitemOpen
  \bibfield  {author} {\bibinfo {author} {\bibfnamefont {A.~P.}\ \bibnamefont
  {Peirce}}, \bibinfo {author} {\bibfnamefont {M.~A.}\ \bibnamefont {Dahleh}},
  \ and\ \bibinfo {author} {\bibfnamefont {H.}~\bibnamefont {Rabitz}},\
  }\href@noop {} {\bibfield  {journal} {\bibinfo  {journal} {Physical Review
  A}\ }\textbf {\bibinfo {volume} {37}},\ \bibinfo {pages} {4950} (\bibinfo
  {year} {1988})}\BibitemShut {NoStop}%
\bibitem [{\citenamefont {Khaneja}\ \emph {et~al.}(2005)\citenamefont
  {Khaneja}, \citenamefont {Reiss}, \citenamefont {Kehlet}, \citenamefont
  {Schulte-Herbrüggen},\ and\ \citenamefont {Glaser}}]{Khaneja2005}%
  \BibitemOpen
  \bibfield  {author} {\bibinfo {author} {\bibfnamefont {N.}~\bibnamefont
  {Khaneja}}, \bibinfo {author} {\bibfnamefont {T.}~\bibnamefont {Reiss}},
  \bibinfo {author} {\bibfnamefont {C.}~\bibnamefont {Kehlet}}, \bibinfo
  {author} {\bibfnamefont {T.}~\bibnamefont {Schulte-Herbrüggen}}, \ and\
  \bibinfo {author} {\bibfnamefont {S.~J.}\ \bibnamefont {Glaser}},\ }\href
  {\doibase https://doi.org/10.1016/j.jmr.2004.11.004} {\bibfield  {journal}
  {\bibinfo  {journal} {Journal of Magnetic Resonance}\ }\textbf {\bibinfo
  {volume} {172}},\ \bibinfo {pages} {296 } (\bibinfo {year}
  {2005})}\BibitemShut {NoStop}%
\bibitem [{\citenamefont {Dolde}\ \emph {et~al.}(2014)\citenamefont {Dolde},
  \citenamefont {Bergholm}, \citenamefont {Wang}, \citenamefont {Jakobi},
  \citenamefont {Naydenov}, \citenamefont {Pezzagna}, \citenamefont {Meijer},
  \citenamefont {Jelezko}, \citenamefont {Neumann}, \citenamefont
  {Schulte-Herbr{\"u}ggen} \emph {et~al.}}]{Dolde2014}%
  \BibitemOpen
  \bibfield  {author} {\bibinfo {author} {\bibfnamefont {F.}~\bibnamefont
  {Dolde}}, \bibinfo {author} {\bibfnamefont {V.}~\bibnamefont {Bergholm}},
  \bibinfo {author} {\bibfnamefont {Y.}~\bibnamefont {Wang}}, \bibinfo {author}
  {\bibfnamefont {I.}~\bibnamefont {Jakobi}}, \bibinfo {author} {\bibfnamefont
  {B.}~\bibnamefont {Naydenov}}, \bibinfo {author} {\bibfnamefont
  {S.}~\bibnamefont {Pezzagna}}, \bibinfo {author} {\bibfnamefont
  {J.}~\bibnamefont {Meijer}}, \bibinfo {author} {\bibfnamefont
  {F.}~\bibnamefont {Jelezko}}, \bibinfo {author} {\bibfnamefont
  {P.}~\bibnamefont {Neumann}}, \bibinfo {author} {\bibfnamefont
  {T.}~\bibnamefont {Schulte-Herbr{\"u}ggen}},  \emph {et~al.},\ }\href@noop {}
  {\bibfield  {journal} {\bibinfo  {journal} {Nature communications}\ }\textbf
  {\bibinfo {volume} {5}},\ \bibinfo {pages} {3371} (\bibinfo {year}
  {2014})}\BibitemShut {NoStop}%
\bibitem [{\citenamefont {Glaser}\ \emph {et~al.}(2015)\citenamefont {Glaser},
  \citenamefont {Boscain}, \citenamefont {Calarco}, \citenamefont {Koch},
  \citenamefont {K{\"o}ckenberger}, \citenamefont {Kosloff}, \citenamefont
  {Kuprov}, \citenamefont {Luy}, \citenamefont {Schirmer}, \citenamefont
  {Schulte-Herbr{\"u}ggen}, \citenamefont {Sugny},\ and\ \citenamefont
  {Wilhelm}}]{Glaser2015}%
  \BibitemOpen
  \bibfield  {author} {\bibinfo {author} {\bibfnamefont {S.~J.}\ \bibnamefont
  {Glaser}}, \bibinfo {author} {\bibfnamefont {U.}~\bibnamefont {Boscain}},
  \bibinfo {author} {\bibfnamefont {T.}~\bibnamefont {Calarco}}, \bibinfo
  {author} {\bibfnamefont {C.~P.}\ \bibnamefont {Koch}}, \bibinfo {author}
  {\bibfnamefont {W.}~\bibnamefont {K{\"o}ckenberger}}, \bibinfo {author}
  {\bibfnamefont {R.}~\bibnamefont {Kosloff}}, \bibinfo {author} {\bibfnamefont
  {I.}~\bibnamefont {Kuprov}}, \bibinfo {author} {\bibfnamefont
  {B.}~\bibnamefont {Luy}}, \bibinfo {author} {\bibfnamefont {S.}~\bibnamefont
  {Schirmer}}, \bibinfo {author} {\bibfnamefont {T.}~\bibnamefont
  {Schulte-Herbr{\"u}ggen}}, \bibinfo {author} {\bibfnamefont {D.}~\bibnamefont
  {Sugny}}, \ and\ \bibinfo {author} {\bibfnamefont {F.~K.}\ \bibnamefont
  {Wilhelm}},\ }\href {\doibase 10.1140/epjd/e2015-60464-1} {\bibfield
  {journal} {\bibinfo  {journal} {The European Physical Journal D}\ }\textbf
  {\bibinfo {volume} {69}},\ \bibinfo {pages} {279} (\bibinfo {year}
  {2015})}\BibitemShut {NoStop}%
\bibitem [{\citenamefont {Wang}\ \emph {et~al.}(2015)\citenamefont {Wang},
  \citenamefont {Allegra}, \citenamefont {Jacobs}, \citenamefont {Lloyd},
  \citenamefont {Lupo},\ and\ \citenamefont {Mohseni}}]{Wang2015}%
  \BibitemOpen
  \bibfield  {author} {\bibinfo {author} {\bibfnamefont {X.}~\bibnamefont
  {Wang}}, \bibinfo {author} {\bibfnamefont {M.}~\bibnamefont {Allegra}},
  \bibinfo {author} {\bibfnamefont {K.}~\bibnamefont {Jacobs}}, \bibinfo
  {author} {\bibfnamefont {S.}~\bibnamefont {Lloyd}}, \bibinfo {author}
  {\bibfnamefont {C.}~\bibnamefont {Lupo}}, \ and\ \bibinfo {author}
  {\bibfnamefont {M.}~\bibnamefont {Mohseni}},\ }\href {\doibase
  10.1103/PhysRevLett.114.170501} {\bibfield  {journal} {\bibinfo  {journal}
  {Phys. Rev. Lett.}\ }\textbf {\bibinfo {volume} {114}},\ \bibinfo {pages}
  {170501} (\bibinfo {year} {2015})}\BibitemShut {NoStop}%
\bibitem [{\citenamefont {Geng}\ \emph {et~al.}(2016)\citenamefont {Geng},
  \citenamefont {Wu}, \citenamefont {Wang}, \citenamefont {Xu}, \citenamefont
  {Shi}, \citenamefont {Xie}, \citenamefont {Rong},\ and\ \citenamefont
  {Du}}]{Geng2016}%
  \BibitemOpen
  \bibfield  {author} {\bibinfo {author} {\bibfnamefont {J.}~\bibnamefont
  {Geng}}, \bibinfo {author} {\bibfnamefont {Y.}~\bibnamefont {Wu}}, \bibinfo
  {author} {\bibfnamefont {X.}~\bibnamefont {Wang}}, \bibinfo {author}
  {\bibfnamefont {K.}~\bibnamefont {Xu}}, \bibinfo {author} {\bibfnamefont
  {F.}~\bibnamefont {Shi}}, \bibinfo {author} {\bibfnamefont {Y.}~\bibnamefont
  {Xie}}, \bibinfo {author} {\bibfnamefont {X.}~\bibnamefont {Rong}}, \ and\
  \bibinfo {author} {\bibfnamefont {J.}~\bibnamefont {Du}},\ }\href {\doibase
  10.1103/PhysRevLett.117.170501} {\bibfield  {journal} {\bibinfo  {journal}
  {Phys. Rev. Lett.}\ }\textbf {\bibinfo {volume} {117}},\ \bibinfo {pages}
  {170501} (\bibinfo {year} {2016})}\BibitemShut {NoStop}%
\bibitem [{\citenamefont {Judson}\ and\ \citenamefont
  {Rabitz}(1992)}]{Judson1992}%
  \BibitemOpen
  \bibfield  {author} {\bibinfo {author} {\bibfnamefont {R.~S.}\ \bibnamefont
  {Judson}}\ and\ \bibinfo {author} {\bibfnamefont {H.}~\bibnamefont
  {Rabitz}},\ }\href {\doibase 10.1103/PhysRevLett.68.1500} {\bibfield
  {journal} {\bibinfo  {journal} {Phys. Rev. Lett.}\ }\textbf {\bibinfo
  {volume} {68}},\ \bibinfo {pages} {1500} (\bibinfo {year}
  {1992})}\BibitemShut {NoStop}%
\bibitem [{\citenamefont {Phan}\ and\ \citenamefont {Rabitz}(1999)}]{Phan1999}%
  \BibitemOpen
  \bibfield  {author} {\bibinfo {author} {\bibfnamefont {M.~Q.}\ \bibnamefont
  {Phan}}\ and\ \bibinfo {author} {\bibfnamefont {H.}~\bibnamefont {Rabitz}},\
  }\href@noop {} {\bibfield  {journal} {\bibinfo  {journal} {The Journal of
  chemical physics}\ }\textbf {\bibinfo {volume} {110}},\ \bibinfo {pages} {34}
  (\bibinfo {year} {1999})}\BibitemShut {NoStop}%
\bibitem [{\citenamefont {Warren}\ \emph {et~al.}(1993)\citenamefont {Warren},
  \citenamefont {Rabitz},\ and\ \citenamefont {Dahleh}}]{Warren1993}%
  \BibitemOpen
  \bibfield  {author} {\bibinfo {author} {\bibfnamefont {W.~S.}\ \bibnamefont
  {Warren}}, \bibinfo {author} {\bibfnamefont {H.}~\bibnamefont {Rabitz}}, \
  and\ \bibinfo {author} {\bibfnamefont {M.}~\bibnamefont {Dahleh}},\ }\href
  {\doibase 10.1126/science.259.5101.1581} {\bibfield  {journal} {\bibinfo
  {journal} {Science}\ }\textbf {\bibinfo {volume} {259}},\ \bibinfo {pages}
  {1581} (\bibinfo {year} {1993})}\BibitemShut {NoStop}%
\bibitem [{\citenamefont {Lee}(2002)}]{Lee2002}%
  \BibitemOpen
  \bibfield  {author} {\bibinfo {author} {\bibfnamefont {J.-S.}\ \bibnamefont
  {Lee}},\ }\href {\doibase https://doi.org/10.1016/S0375-9601(02)01479-2}
  {\bibfield  {journal} {\bibinfo  {journal} {Physics Letters A}\ }\textbf
  {\bibinfo {volume} {305}},\ \bibinfo {pages} {349 } (\bibinfo {year}
  {2002})}\BibitemShut {NoStop}%
\bibitem [{\citenamefont {Sun}\ \emph {et~al.}(2014)\citenamefont {Sun},
  \citenamefont {Pelczer}, \citenamefont {Riviello}, \citenamefont {Wu},\ and\
  \citenamefont {Rabitz}}]{Sun2014}%
  \BibitemOpen
  \bibfield  {author} {\bibinfo {author} {\bibfnamefont {Q.}~\bibnamefont
  {Sun}}, \bibinfo {author} {\bibfnamefont {I.}~\bibnamefont {Pelczer}},
  \bibinfo {author} {\bibfnamefont {G.}~\bibnamefont {Riviello}}, \bibinfo
  {author} {\bibfnamefont {R.-B.}\ \bibnamefont {Wu}}, \ and\ \bibinfo {author}
  {\bibfnamefont {H.}~\bibnamefont {Rabitz}},\ }\href {\doibase
  10.1103/PhysRevA.89.033413} {\bibfield  {journal} {\bibinfo  {journal} {Phys.
  Rev. A}\ }\textbf {\bibinfo {volume} {89}},\ \bibinfo {pages} {033413}
  (\bibinfo {year} {2014})}\BibitemShut {NoStop}%
\bibitem [{\citenamefont {Sun}\ \emph {et~al.}(2015)\citenamefont {Sun},
  \citenamefont {Pelczer}, \citenamefont {Riviello}, \citenamefont {Wu},\ and\
  \citenamefont {Rabitz}}]{Sun2015}%
  \BibitemOpen
  \bibfield  {author} {\bibinfo {author} {\bibfnamefont {Q.}~\bibnamefont
  {Sun}}, \bibinfo {author} {\bibfnamefont {I.}~\bibnamefont {Pelczer}},
  \bibinfo {author} {\bibfnamefont {G.}~\bibnamefont {Riviello}}, \bibinfo
  {author} {\bibfnamefont {R.-B.}\ \bibnamefont {Wu}}, \ and\ \bibinfo {author}
  {\bibfnamefont {H.}~\bibnamefont {Rabitz}},\ }\href {\doibase
  10.1103/PhysRevA.91.043412} {\bibfield  {journal} {\bibinfo  {journal} {Phys.
  Rev. A}\ }\textbf {\bibinfo {volume} {91}},\ \bibinfo {pages} {043412}
  (\bibinfo {year} {2015})}\BibitemShut {NoStop}%
\bibitem [{\citenamefont {Sun}\ \emph {et~al.}(2017)\citenamefont {Sun},
  \citenamefont {Wu},\ and\ \citenamefont {Rabitz}}]{Sun2017}%
  \BibitemOpen
  \bibfield  {author} {\bibinfo {author} {\bibfnamefont {Q.}~\bibnamefont
  {Sun}}, \bibinfo {author} {\bibfnamefont {R.-B.}\ \bibnamefont {Wu}}, \ and\
  \bibinfo {author} {\bibfnamefont {H.}~\bibnamefont {Rabitz}},\ }\href
  {\doibase 10.1103/PhysRevA.95.032319} {\bibfield  {journal} {\bibinfo
  {journal} {Phys. Rev. A}\ }\textbf {\bibinfo {volume} {95}},\ \bibinfo
  {pages} {032319} (\bibinfo {year} {2017})}\BibitemShut {NoStop}%
\bibitem [{\citenamefont {Li}\ \emph {et~al.}(2017)\citenamefont {Li},
  \citenamefont {Yang}, \citenamefont {Peng},\ and\ \citenamefont
  {Sun}}]{Li2017}%
  \BibitemOpen
  \bibfield  {author} {\bibinfo {author} {\bibfnamefont {J.}~\bibnamefont
  {Li}}, \bibinfo {author} {\bibfnamefont {X.}~\bibnamefont {Yang}}, \bibinfo
  {author} {\bibfnamefont {X.}~\bibnamefont {Peng}}, \ and\ \bibinfo {author}
  {\bibfnamefont {C.-P.}\ \bibnamefont {Sun}},\ }\href@noop {} {\bibfield
  {journal} {\bibinfo  {journal} {Physical review letters}\ }\textbf {\bibinfo
  {volume} {118}},\ \bibinfo {pages} {150503} (\bibinfo {year}
  {2017})}\BibitemShut {NoStop}%
\bibitem [{\citenamefont {Lu}\ \emph {et~al.}(2017)\citenamefont {Lu},
  \citenamefont {Li}, \citenamefont {Li}, \citenamefont {Katiyar},
  \citenamefont {Park}, \citenamefont {Feng}, \citenamefont {Xin},
  \citenamefont {Li}, \citenamefont {Long}, \citenamefont {Brodutch},
  \citenamefont {Baugh}, \citenamefont {Zeng},\ and\ \citenamefont
  {Laflamme}}]{Lu2017}%
  \BibitemOpen
  \bibfield  {author} {\bibinfo {author} {\bibfnamefont {D.}~\bibnamefont
  {Lu}}, \bibinfo {author} {\bibfnamefont {K.}~\bibnamefont {Li}}, \bibinfo
  {author} {\bibfnamefont {J.}~\bibnamefont {Li}}, \bibinfo {author}
  {\bibfnamefont {H.}~\bibnamefont {Katiyar}}, \bibinfo {author} {\bibfnamefont
  {A.~J.}\ \bibnamefont {Park}}, \bibinfo {author} {\bibfnamefont
  {G.}~\bibnamefont {Feng}}, \bibinfo {author} {\bibfnamefont {T.}~\bibnamefont
  {Xin}}, \bibinfo {author} {\bibfnamefont {H.}~\bibnamefont {Li}}, \bibinfo
  {author} {\bibfnamefont {G.}~\bibnamefont {Long}}, \bibinfo {author}
  {\bibfnamefont {A.}~\bibnamefont {Brodutch}}, \bibinfo {author}
  {\bibfnamefont {J.}~\bibnamefont {Baugh}}, \bibinfo {author} {\bibfnamefont
  {B.}~\bibnamefont {Zeng}}, \ and\ \bibinfo {author} {\bibfnamefont
  {R.}~\bibnamefont {Laflamme}},\ }\href@noop {} {\bibfield  {journal}
  {\bibinfo  {journal} {npj Quantum Information}\ }\textbf {\bibinfo {volume}
  {3}},\ \bibinfo {pages} {45} (\bibinfo {year} {2017})}\BibitemShut {NoStop}%
\bibitem [{\citenamefont {Egger}\ and\ \citenamefont
  {Wilhelm}(2014)}]{egger2014adaptive}%
  \BibitemOpen
  \bibfield  {author} {\bibinfo {author} {\bibfnamefont {D.}~\bibnamefont
  {Egger}}\ and\ \bibinfo {author} {\bibfnamefont {F.~K.}\ \bibnamefont
  {Wilhelm}},\ }\href@noop {} {\bibfield  {journal} {\bibinfo  {journal} {Phys.
  Rev. Lett.}\ }\textbf {\bibinfo {volume} {112}},\ \bibinfo {pages} {240503}
  (\bibinfo {year} {2014})}\BibitemShut {NoStop}%
\bibitem [{\citenamefont {Dive}\ \emph {et~al.}(2017)\citenamefont {Dive},
  \citenamefont {Pitchford}, \citenamefont {Mintert},\ and\ \citenamefont
  {Burgarth}}]{dive2017situ}%
  \BibitemOpen
  \bibfield  {author} {\bibinfo {author} {\bibfnamefont {B.}~\bibnamefont
  {Dive}}, \bibinfo {author} {\bibfnamefont {A.}~\bibnamefont {Pitchford}},
  \bibinfo {author} {\bibfnamefont {F.}~\bibnamefont {Mintert}}, \ and\
  \bibinfo {author} {\bibfnamefont {D.}~\bibnamefont {Burgarth}},\ }\href@noop
  {} {\bibfield  {journal} {\bibinfo  {journal} {arXiv preprint
  arXiv:1701.01723}\ } (\bibinfo {year} {2017})}\BibitemShut {NoStop}%
\bibitem [{\citenamefont {Wu}\ \emph {et~al.}(2018)\citenamefont {Wu},
  \citenamefont {Chu}, \citenamefont {Owens},\ and\ \citenamefont
  {Rabitz}}]{wu2018data}%
  \BibitemOpen
  \bibfield  {author} {\bibinfo {author} {\bibfnamefont {R.-B.}\ \bibnamefont
  {Wu}}, \bibinfo {author} {\bibfnamefont {B.}~\bibnamefont {Chu}}, \bibinfo
  {author} {\bibfnamefont {D.~H.}\ \bibnamefont {Owens}}, \ and\ \bibinfo
  {author} {\bibfnamefont {H.}~\bibnamefont {Rabitz}},\ }\href@noop {}
  {\bibfield  {journal} {\bibinfo  {journal} {Physical Review A}\ }\textbf
  {\bibinfo {volume} {97}},\ \bibinfo {pages} {042122} (\bibinfo {year}
  {2018})}\BibitemShut {NoStop}%
\bibitem [{Note1()}]{Note1}%
  \BibitemOpen
  Note1,\ \href@noop {} {}\bibinfo {note} {In the NMR community this result is
  sometimes referred to as a pseudo-pure state}\BibitemShut {NoStop}%
\bibitem [{\citenamefont {Peng}\ \emph {et~al.}(2001)\citenamefont {Peng},
  \citenamefont {Zhu}, \citenamefont {Fang}, \citenamefont {Feng},
  \citenamefont {Gao}, \citenamefont {Yang},\ and\ \citenamefont
  {Liu}}]{Peng2001}%
  \BibitemOpen
  \bibfield  {author} {\bibinfo {author} {\bibfnamefont {X.}~\bibnamefont
  {Peng}}, \bibinfo {author} {\bibfnamefont {X.}~\bibnamefont {Zhu}}, \bibinfo
  {author} {\bibfnamefont {X.}~\bibnamefont {Fang}}, \bibinfo {author}
  {\bibfnamefont {M.}~\bibnamefont {Feng}}, \bibinfo {author} {\bibfnamefont
  {K.}~\bibnamefont {Gao}}, \bibinfo {author} {\bibfnamefont {X.}~\bibnamefont
  {Yang}}, \ and\ \bibinfo {author} {\bibfnamefont {M.}~\bibnamefont {Liu}},\
  }\href {\doibase https://doi.org/10.1016/S0009-2614(01)00421-3} {\bibfield
  {journal} {\bibinfo  {journal} {Chemical Physics Letters}\ }\textbf {\bibinfo
  {volume} {340}},\ \bibinfo {pages} {509 } (\bibinfo {year}
  {2001})}\BibitemShut {NoStop}%
\bibitem [{\citenamefont {Arenz}\ and\ \citenamefont
  {Rabitz}(2018)}]{Arenz2018}%
  \BibitemOpen
  \bibfield  {author} {\bibinfo {author} {\bibfnamefont {C.}~\bibnamefont
  {Arenz}}\ and\ \bibinfo {author} {\bibfnamefont {H.}~\bibnamefont {Rabitz}},\
  }\href {\doibase 10.1103/PhysRevLett.120.220503} {\bibfield  {journal}
  {\bibinfo  {journal} {Phys. Rev. Lett.}\ }\textbf {\bibinfo {volume} {120}},\
  \bibinfo {pages} {220503} (\bibinfo {year} {2018})}\BibitemShut {NoStop}%
\bibitem [{\citenamefont {Hegerfeldt}(2013)}]{Hegerfeldt2013}%
  \BibitemOpen
  \bibfield  {author} {\bibinfo {author} {\bibfnamefont {G.~C.}\ \bibnamefont
  {Hegerfeldt}},\ }\href@noop {} {\bibfield  {journal} {\bibinfo  {journal}
  {Physical review letters}\ }\textbf {\bibinfo {volume} {111}},\ \bibinfo
  {pages} {260501} (\bibinfo {year} {2013})}\BibitemShut {NoStop}%
\bibitem [{\citenamefont {Khaneja}\ \emph {et~al.}(2001)\citenamefont
  {Khaneja}, \citenamefont {Brockett},\ and\ \citenamefont
  {Glaser}}]{Khaneja2001}%
  \BibitemOpen
  \bibfield  {author} {\bibinfo {author} {\bibfnamefont {N.}~\bibnamefont
  {Khaneja}}, \bibinfo {author} {\bibfnamefont {R.}~\bibnamefont {Brockett}}, \
  and\ \bibinfo {author} {\bibfnamefont {S.~J.}\ \bibnamefont {Glaser}},\
  }\href {\doibase 10.1103/PhysRevA.63.032308} {\bibfield  {journal} {\bibinfo
  {journal} {Phys. Rev. A}\ }\textbf {\bibinfo {volume} {63}},\ \bibinfo
  {pages} {032308} (\bibinfo {year} {2001})}\BibitemShut {NoStop}%
\bibitem [{\citenamefont {Khaneja}\ \emph {et~al.}(2002)\citenamefont
  {Khaneja}, \citenamefont {Glaser},\ and\ \citenamefont
  {Brockett}}]{Khaneja2002}%
  \BibitemOpen
  \bibfield  {author} {\bibinfo {author} {\bibfnamefont {N.}~\bibnamefont
  {Khaneja}}, \bibinfo {author} {\bibfnamefont {S.~J.}\ \bibnamefont {Glaser}},
  \ and\ \bibinfo {author} {\bibfnamefont {R.}~\bibnamefont {Brockett}},\
  }\href@noop {} {\bibfield  {journal} {\bibinfo  {journal} {Physical Review
  A}\ }\textbf {\bibinfo {volume} {65}},\ \bibinfo {pages} {032301} (\bibinfo
  {year} {2002})}\BibitemShut {NoStop}%
\bibitem [{\citenamefont {Carlini}\ \emph {et~al.}(2006)\citenamefont
  {Carlini}, \citenamefont {Hosoya}, \citenamefont {Koike},\ and\ \citenamefont
  {Okudaira}}]{Carlini2006}%
  \BibitemOpen
  \bibfield  {author} {\bibinfo {author} {\bibfnamefont {A.}~\bibnamefont
  {Carlini}}, \bibinfo {author} {\bibfnamefont {A.}~\bibnamefont {Hosoya}},
  \bibinfo {author} {\bibfnamefont {T.}~\bibnamefont {Koike}}, \ and\ \bibinfo
  {author} {\bibfnamefont {Y.}~\bibnamefont {Okudaira}},\ }\href@noop {}
  {\bibfield  {journal} {\bibinfo  {journal} {Physical review letters}\
  }\textbf {\bibinfo {volume} {96}},\ \bibinfo {pages} {060503} (\bibinfo
  {year} {2006})}\BibitemShut {NoStop}%
\bibitem [{\citenamefont {Chen}\ \emph {et~al.}(2015)\citenamefont {Chen},
  \citenamefont {Wu}, \citenamefont {Zhang},\ and\ \citenamefont
  {Rabitz}}]{Chen2015}%
  \BibitemOpen
  \bibfield  {author} {\bibinfo {author} {\bibfnamefont {Q.-M.}\ \bibnamefont
  {Chen}}, \bibinfo {author} {\bibfnamefont {R.-B.}\ \bibnamefont {Wu}},
  \bibinfo {author} {\bibfnamefont {T.-M.}\ \bibnamefont {Zhang}}, \ and\
  \bibinfo {author} {\bibfnamefont {H.}~\bibnamefont {Rabitz}},\ }\href
  {\doibase 10.1103/PhysRevA.92.063415} {\bibfield  {journal} {\bibinfo
  {journal} {Phys. Rev. A}\ }\textbf {\bibinfo {volume} {92}},\ \bibinfo
  {pages} {063415} (\bibinfo {year} {2015})}\BibitemShut {NoStop}%
\bibitem [{\citenamefont {Boyd}\ and\ \citenamefont
  {Vandenberghe}(2004)}]{boyd2004convex}%
  \BibitemOpen
  \bibfield  {author} {\bibinfo {author} {\bibfnamefont {S.}~\bibnamefont
  {Boyd}}\ and\ \bibinfo {author} {\bibfnamefont {L.}~\bibnamefont
  {Vandenberghe}},\ }\href@noop {} {\emph {\bibinfo {title} {Convex
  optimization}}}\ (\bibinfo  {publisher} {Cambridge university press},\
  \bibinfo {year} {2004})\BibitemShut {NoStop}%
\bibitem [{Note2()}]{Note2}%
  \BibitemOpen
  Note2,\ \href@noop {} {}\bibinfo {note} {Since $J_{H}=0.999$ is fixed we
  remark that the present procedure seeks a single point along the Pareto
  front.}\BibitemShut {Stop}%
\bibitem [{\citenamefont {Anklin}()}]{PrivateComBruk}%
  \BibitemOpen
  \bibfield  {author} {\bibinfo {author} {\bibfnamefont {C.}~\bibnamefont
  {Anklin}},\ }\href@noop {} {\bibinfo  {journal} {Bruker-Biospin, personal
  communication}\ }\BibitemShut {NoStop}%
\bibitem [{\citenamefont {Keeler}(2003)}]{sorensen2006james}%
  \BibitemOpen
\bibfield  {journal} {  }\bibfield  {author} {\bibinfo {author} {\bibfnamefont
  {J.}~\bibnamefont {Keeler}},\ }\href@noop {} {\bibfield  {journal} {\bibinfo
  {journal} {Understanding NMR Spectroscopy}\ ,\ \bibinfo {pages} {Wiley}}
  (\bibinfo {year} {2003})}\BibitemShut {NoStop}%
\bibitem [{\citenamefont {Vandersypen}\ and\ \citenamefont
  {Chuang}(2005)}]{Vandersypen2005}%
  \BibitemOpen
  \bibfield  {author} {\bibinfo {author} {\bibfnamefont {L.~M.~K.}\
  \bibnamefont {Vandersypen}}\ and\ \bibinfo {author} {\bibfnamefont {I.~L.}\
  \bibnamefont {Chuang}},\ }\href {\doibase 10.1103/RevModPhys.76.1037}
  {\bibfield  {journal} {\bibinfo  {journal} {Rev. Mod. Phys.}\ }\textbf
  {\bibinfo {volume} {76}},\ \bibinfo {pages} {1037} (\bibinfo {year}
  {2005})}\BibitemShut {NoStop}%
\end{thebibliography}%
\end{document}